\documentclass{amsart}

\usepackage[T1]{fontenc}

\usepackage{amsmath,amssymb}

\usepackage{algorithm,algpseudocode}

\usepackage{ifpdf,graphicx}

\usepackage{soul,color}

\DeclareMathOperator{\rand}{rand}

\begin{document}

\title[Full restoration of visual encrypted color images]{%
  Full restoration of\\ visual encrypted color images}

\author{Simeon Persson}
\curraddr{Blekinge Institute of Technology, Undegraduate Student}
\email{}

\author{Kristian V\'arnai}
\curraddr{Blekinge Institute of Technology, Undegraduate Student}
\email{}



\maketitle

\section{Abstract}

While strictly black and white images have been the basis for visual 
cryptography, there has been a lack of an easily implemented format for
colour images. This paper establishes a simple, yet secure way of implementing
visual cryptography with colour, assuming a binary data representation.

\section{Introduction}

Traditional methods for visual cryptography have been established, are
consistent and easily understood. Unfortunately, these methods exist for
black and white pictures only, leaving the encryption of colour images
wanting. While there are a handful of attempts at bringing colour to visual
cryptography, it is still an open field with implementations of varying
efficiency. In this paper, we will establish some basic standards for
encrypting colour pictures, as well as a simple, yet efficient method for
encryption based upon those rules.

While black and white pictures are fairly easy to work with due to their
simple nature, colour pictures contain much more information and possibly
details that may not be lost in the process. This leads to a need for the
static that normally appears in the decryption of black and white pictures to
have to be absent[1]. However, a partial reconstruction of the picture (such as
having less than all the necessary parts for full restoration) may not hint
at what the final image is meant to be.

In addition to the standard of security for the traditional black and white
pictures, we set the following points as mandatory for encryption of colour
images.
\begin{itemize}
\item Full restoration upon decryption.
\item No indication as to the original image, whether by eye or any other
  method, when combining a subset of all available parts.
\item Usability for any type of image, whether that image contains a mixture
  of colours, is black and white or is simply one single colour.
\item Destruction of intermediary steps in the encryption process.
\end{itemize}
Our method of encryption is, due to the process of creating images within
visual cryptography and the expanse of computer use, based around the RGB
colour model for computers and other, similar devices. This does not prevent
implementation of this process with any other model as long as the
information is stored as bits.

\section{Encryption}

The encryption is fairly straightforward and can be easily understood as well
as implemented.  Although it is easy to handle, it is efficient, provides the
necessary security and fulfills every point previously stated. For the
process, the thought was to work on the bitwise level that represents the
colours themselves; in this case, the RGB values are used.  As each pixel is
processed, two random values are generated; the first one is compared to the
RGB value of the current pixel and it is then separated into two values: an
RGB value with the bits that were set in both the original as well as the
first random value, and another one with the set bits left over from the
original.

Next, the second random value is compared to the two new results from the
previous step. If both values are not set while the bit at the same position
in the random value is, those bits for the values from the previous step are
set.  These steps are repeated for every pixel and then the encryption is
finished. Decryption is easily done via a bitwise XOR of the RGB values of
the two resulting pictures and we effectively have a One-Time Pad
implementation on the colour values of an image. The random values from the
process of encryption are discarded alongside any other values we might have
produced.

The encryption algorithm will ``split'' the original image into two so 
called \emph{shadow images}.  Let $p$ denote one pixel in the original 
image, and let $s_1$ and $s_2$ denote the corresponding pixels in the shadow 
images, respectively.  Here $p$, $s_1$, and $s_2$ are vectors, representing
the channels used, e.g. red, green and blue in the RGB colour model. The 
calculations during the encryption are carried out both bitwise and channel-wise.
\begin{algorithm}
  \caption{Visual encryption}
  \begin{algorithmic}[1]
    \ForAll{pixel $p$ in the original image}
    \State $r_1 \gets \rand$
    \State $r_2 \gets \rand$
    \State $s_1 \gets (p \land r_1) \lor (\neg p \land r_2)$
    \Comment{Corresponding pixel in shadow 1}
    \State $s_2 \gets p \oplus s_1$
    \Comment{Corresponding pixel in shadow 2}
    \EndFor
  \end{algorithmic}
\end{algorithm}

\begin{figure}
  \centering
  \ifpdf
    \includegraphics[width=4cm]{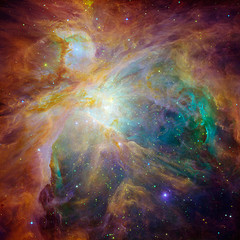}
  \fi
  \caption{Original image}\label{fig:ex1}
\end{figure}

\begin{figure}
  \centering
  \ifpdf
    \includegraphics[width=4cm]{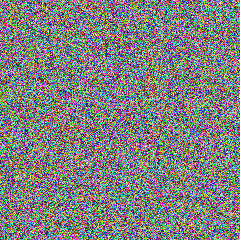}\qquad
    \includegraphics[width=4cm]{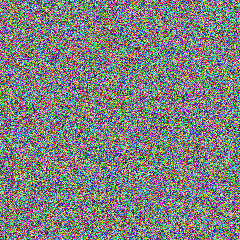}
  \fi
  \caption{Shadow images}\label{fig:ex2}
\end{figure}

We illustrate the encryption with an example.  Assume that we want
to encrypt the image in Figure~\ref{fig:ex1}.  Then if we apply the
algorithm, it may result in the two shadow images in
Figure~\ref{fig:ex2}---remember that the algorithm is probabilistic.
To restore the original image, we compute $s_1 \oplus s_2$ for each pixel.
The result will give us back the original image, without any loss of
quality, that is, Figure~\ref{fig:ex1}.

\section{Validation of the Security}

As with both visual cryptography and the one time pad, this method offers
complete security as there is neither a repetition to be found, nor is a
brute force attack possible as every possible result within the picture's
resolution will show up. This method does not only fulfill the standards we
previously set, it even leaves the resolution of the original image
intact. Of course, it should be noted that for this encryption to work to its
full potential, the results must be saved as a lossless image type. In the
event that one would wish to separate the original into more than two
pictures, reapplying this process to the results until a satisfactory amount
is reached is all one needs to do.

It could be argued that all this encryption would need is random values
generated and applied with a bitwise XOR to the original image, leaving the
randomly generated sequence as one of the resulting images and the result of
the bitwise XOR as the other. While this could be done, let us take a look at
the absolute worst case scenario, disregarding the possibilities that an
attacker knows the original or has access to all parts of the picture. The
scenario in mind would be in the highly unlikely event that an attacker would
have an intimate enough knowledge of the encryption process to know exactly
how it is implemented as well as knowing exactly which random values were
generated.

If the method of encryption would be nothing more than the simplified version
suggested, then having the result of the bitwise XOR in your possession would
be enough to get the original as being able to predict the exact
pseudo-random values would mean that you effectively have the key. This is
not the case for the method we propose as depending on the random values and
how they interact with the original data, only some bits could possibly be
decided, as shown in table~\ref{tab:values}.  There are four possible cases
  for the pair $(r_1, r_2)$.  Each case can be divided into two subcases,
  where each corresponds to the possible value $q$ of $p$.  Without loss of
  generality, we consider the cases on bit-level.

\begin{table}
  \centering
  \begin{tabular}{*{5}{c}}
    \hline
    Case & $r_1$ & $r_2$ & $q$ & $p$ \\
    \hline
    1 & 0 & 0 & 0 & -- \\
    2 & 0 & 0 & 1 & 1  \\
    3 & 0 & 1 & 0 & 1  \\
    4 & 0 & 1 & 1 & -- \\
    5 & 1 & 0 & 0 & -- \\
    6 & 1 & 0 & 1 & 1  \\
    7 & 1 & 1 & 0 & 1  \\
    8 & 1 & 1 & 1 & -- \\
    \hline
  \end{tabular}
  \caption{}\label{tab:values}
\end{table}

As seen, even in the case of knowing the generated pseudo-random values, half
the possible combinations lack a value, thus mitigating such an attack. The
method of encryption presented here could also be applied to the one time pad
due to the likeness of them, so providing the extra bit of security by
mitigating this most unlikely of scenarios.  The reasoning for the received
original value presented in the above table is shown with simple boolean
algebra. The equations used are the ones presented in the pseudocode, also
presented here for ease of following. Keep in mind that we do not know which
one of the values we have and thus not what the other one could be.

\paragraph{\bfseries Case 1 and 2}
Here is $s_1 = 0$ and $s_2 = p \oplus 0$.  While we know for sure that at
least one of the values must be $0$, the other could be either possible
value. This means that if our available result ($q$) is $0$, we have no way of
knowing which value we have in our possession, which in turn means that
$s_1$ and $s_2$ are equal in terms of possibility. Thus, since the original value 
could be either for $s_1$, it remains undetermined. If, on the other hand, our 
available value is $1$, then we can logically conclude that it is $s_2$ that we are 
dealing with and we only have to complete the following equation: $1 = p \oplus 
0$ and therefore $p = 1$.

\paragraph{\bfseries Case 3 and 4}
Here is $s_1 = \neg p$ and $s_2 = 1$.  As per the same logic as in the
previous one: if the available result is the same as the one we know must
exist (in this case one), then we can not tell which equation it is we
have. If it is the opposing value, then we can in this case rule out $s_2$
and calculate the equation: $0 = \neg p$ and hence $p = 1$.

\paragraph{\bfseries Case 5 and 6}
Here is $s_1 = p$ and $s_2 = 0$.  Following the previous logical steps gives
us an unknown for the available result of $0$, but the following result
for $1$ is $p = 1$.

\paragraph{\bfseries Case 7 and 8}
Here is $s_1 = 1$ and $s_2 = p \oplus 1$.  Once again we can not say for sure
which case is before us if our available result is the same as the one we
know with certainty to exist, but in the case of a differing result, we know
the following: $0 = p \oplus 1$ and $p = 1$.

\section{Acknowledgements}
We would like to thank Robert Nyqvist for:
\begin{itemize}
\item Supplying the necessary information
\item Helping with the writing of this paper
\end{itemize}

\end{document}